\documentstyle[pra,aps,epsfig]{revtex}
\begin{document}
\draft
\title{Identification of Beutler-Fano formula in  eigenphase shifts
  and eigentime delays near a resonance}
\author{Chun-Woo Lee}
\address{Department of Chemistry, Ajou University, 5 Wonchun Dong,
  Suwon 442-749, KOREA} 
\maketitle
\begin{abstract}
Eigenphase shifts and eigentime delays near a resonance for
a system of one discrete state and two continua are shown to be
functionals of the Beutler-Fano  formula using appropriate
dimensionless energy units and line profile indices. Parameters
responsible for the avoided crossing of eigenphase shifts and 
eigentime
delays are identified. Similarly, parameters responsible for the
eigentime delays due to a frame change are identified. With the help of
new parameters, analogy with the spin
model is pursued for the $S$ matrix and time delay matrix
$Q$. The $S$
matrix is found to be put into $\exp [ i(a+b \sigma \cdot \hat{n} )]$.
The time delay matrix $Q$ is shown to be given as $Q =
\frac{1}{2} \tau_r (1+ \vec{P_a} \cdot \vec{\sigma} + \vec{P_f} \cdot
\vec{\sigma} )$ where the first term is the
time delay due to 
resonance, the second term is the one due to avoided crossing
interaction and the last term is the one due 
to a frame change. It is found that $P_a^2 + P_f^2$
equals unity.  
\end{abstract}

\section{Introduction}

Resonances observed in the energy-dependence of atomic and molecular
processes often correspond to an autoionization or predissociation
observed in photo-absorption, i.e., to the excitation of a quasi-bound
state in a continuum region\cite{1,2}.
Their spectra are
characterized by rich profiles  providing a wealth of
information on the atomic and molecular dissociation or ionization
processes. Beutler's early observation of rare-gas levels auto-ionizing
into a single continuum with angular-momentum conservation was
reproduced by Fano through the formula\cite{3}:
\begin{equation}
\sigma (\epsilon ) = \sigma _{0} \frac{(\epsilon -q)^2}{1+\epsilon
^2 } , \label{fb}
\end{equation}
where $\epsilon$ is the reduced  energy defined by $2(E-E_{0})
/\Gamma$;
$E_{0}$ and $\hbar / \Gamma$ represent the resonance energy and
the mean life time of the quasi-bound state, respectively;
$\sigma_{0}$ represents the photo-fragmentation cross section to
the continuum state; $q$ is an index that characterizes the line
profile (here $-q$ is employed instead of the usual $+q$).
Extension of Beutler-Fano formula to
the dissociation into multi-channels was obtained by
Combet-Farnoux\cite{4}. 
The energy dependence of  $S$ matrix near an isolated resonance for
such predissociation or autoionization into
multi-channels
was obtained long ago and well known\cite{5}. By diagonalizing $S$
matrix, eigenphase shifts $\delta_{i}$ ($i$=1,2,...$n$) are obtained
from its eigenvalues as 
\begin{equation}
S = Ue^{2i\delta}\tilde{U} , \label{eigenphase}
\end{equation}
and  utilized frequently as a tool for analyzing the
resonances\cite{6,7}. Eigenphase shifts and corresponding
eigenchannels are also 
extensively used in various forms in 
Multichannel Quantum Defect Theory which is regarded as one of the
most general and powerful theories of resonance\cite{1}.  

In contrast to photo-fragmentation cross
sections and $S$ matrix, the analytical
formulas for eigenphase shifts in multichannel processes
are not known. According to the numerical studies, eigenphase
shifts in the multichannel system are known as showing complicated
behaviors near a resonance due 
to the avoided crossing between curves of eigenphase shifts along the
energy\cite{6,8}. Such a phenomenon of the avoided crossing along the
energy 
is not conspicuous in case of the formulas for photo-dissociation cross
sections and $S$ matrices. We report here the detailed studies on the
behaviors of 
eigenphase shifts and also times delayed by a collision\cite{9,10} for the
system of one discrete state and two continua, which turn out to be
described by functionals of the Beutler-Fano formula.
Their behaviors in the system of one discrete state and one
continuum are already studied\cite{3}. If one more continuum is added
to that system, there enter new
phenomena of avoided crossing between  curves of eigenphase shifts
and eigentimes delayed and of times delayed due to a frame change along
the energy besides  the resonance behavior.
The avoided crossing interactions
between eigenphase shifts along the energy always take place near 
a resonance if
they are not excluded for symmetry reasons, as we will see in the later 
section. 
The addition of more continua may simply introduce more avoided 
crossings among the newly added eigenphase shifts. In this sense,
the current system of one discrete state and
two continua may serve as a prototype system 
for the study of  the effects of avoid crossings between eigenphase
shifts and times delayed due to the frame change on the resonance
phenomena. 

The import of the current work may be in the findings of the new 
parameters for the avoided crossing interaction, where $S$ matrix 
can be put into a form of $e^{iH}$ with $H$ hermitian and the time delay
matrix (or life time matrix as called by Smith\cite{10}) can be put
into a form of $1+\vec{P} \cdot \vec{\sigma}$ where $\vec{P}$ is a
polarization vector 
and $\vec{\sigma}$ is the Pauli matrix. 
The analogy with the spin system under a magnetic field is exploited
to find further properties of $S$ matrix and time delay matrix.

Section \ref{sec:known} describes the known properties of eigenphase
shifts. Section \ref{sec:eqn_eps} describes the 
form of $S$ matrix 
near a resonance and the equation for eigenphase shifts. It also
discusses some special cases and general characteristics of eigenphase
shifts and eigentime delays. Sec. \ref{sec:eps_one_discrete_two_continua}
obtains the formulas of eigenphase shifts as functionals of the
Beutler-Fano formula for the system of one discrete state and two
continua. Sec. \ref{sec:etd} obtains the
formulas of eigentime delays as functionals of 
Beutler-Fano formula. 

\section{Known properties of eigenphase shifts}
\label{sec:known}

It is well known that , for an isolated resonance in single channel
scattering, the energy dependence of a phase shift, $\delta (E)$, is
described by the following formula\cite{1}
\begin{equation}
\tan [\delta (E) - \delta ^{0}] = \frac{\Gamma}{2(E_{0} -E)}
= - \frac{1}{\epsilon} , \label{tan}
\end{equation}
where $\delta^{0}$, $E_{0}$ and $\Gamma$ denote the background
phase shift, the energy and the width of the
resonance, respectively. The second equality in Eq. (\ref{tan}) defines 
the dimensionless energy. It can be deduced from Eq. (\ref{tan})  that
$\delta (E)$  
increases  slowly  except for the narrow
energy range  
around $E_{0}$ where its value  undergoes a rapid change and 
is eventually increased by $\pi$ as energy varies from $-\infty$ to
$\infty$.  Due to the property of tangent function, 
$\delta (E)$  can be determined only up to $n \pi$ ($n$ integer). It is
usually taken to be $\delta^{0}$ at $E \rightarrow -\infty$. Then,
it becomes $\delta^{0} + \pi$ at $E \rightarrow \infty$. 
In the absence of  resonance, $\delta$ is equal to $\delta^{0}$.
The background
phase shift $\delta^{0}$ is a slowly varying function of energy
and usually taken 
to be constant as a good approximation. Thus, the
increase of $\delta (E)$ around $E_{0}$ solely comes from
resonance. 
Let us 
denote the difference between $\delta (E)$ and $\delta^{0}$ by
$\delta_r (E)$.  With this notation, Eq. (\ref{tan}) becomes $\tan
\delta_r (E)$ = $\tan \bar{\delta}_r 
(\epsilon )$ = $-1/ \epsilon $, and
with the mentioned phase convention of setting $\delta (E)$ to 
$\delta^{0}$ at $E = -\infty$, it can be transformed as follows
\begin{equation}
e^{i \bar{\delta}_r (\epsilon )} = \frac{-\epsilon +i}{\sqrt{\epsilon
^2 +1}} . \label{delta_single}
\end{equation}
 
Macek\cite{8} generalized Eq. (\ref{tan}) to multichannel processes as
follows
\begin{equation}
2(E-E_{0} ) = \sum_{k=1}^n \Gamma_k \cot [\delta_k^{0} -
\delta_m (E)], ~~ m=1,\ldots,n , \label{tan_mc}
\end{equation}
where 
$\delta_m (E)$ 
is the eigenphase shift defined in  Eq. (\ref{eigenphase}),
$\delta_k^{0}$ is the background phase shifts in channel $k$, and
$\Gamma_k$ is the partial decay width to channel $k$.

Hazi\cite{11} also showed that the eigenphase sum, which will be
denoted as $\delta_{\Sigma}
(E)$, satisfies the 
``single-channel'' formula
\begin{equation}
\delta_{\Sigma} (E) \equiv \sum_{k=1}^n \delta_k (E) =
\delta_{\Sigma}^{0} + 
\tan^{-1} 
\frac{\Gamma}{2(E_{0} - E)}= \delta_{\Sigma}^{0} - \tan^{-1}
\frac{1}{\epsilon} , \label{Hazi}
\end{equation}
where $\Gamma$ is the total width of the resonance, i.e. $\Gamma =
\sum_{m=1}^n 
\Gamma_m$,  and
$\delta_{\Sigma}^{0}$ is the sum of the background eigenphase,
i.e. $\delta_{\Sigma}^{0} = \sum_{m=1}^n \delta_m^{0}$.
The reduced energy $\epsilon$ is the one defined in Eq. (\ref{tan})
except that
$\Gamma$  now denotes the
total width.

\section{Equations for eigenphase shifts in the neighborhood of
resonance}
\label{sec:eqn_eps}

The form of $S$ matrix in the neighborhood of an
isolated resonance in multichannel processes is well known and
repeatedly derived in the past using 
various resonance theories\cite{2,5}. Though its form is basically
same for all resonance theories, its detailed expressions may look
differently from theory to theory.  Here, Fano's configuration
interaction theory is adopted for the treatment of resonance\cite{3}.
In the configuration interaction theory,  an isolated resonance in 
multichannel system is treated as regarding the system as composed of
one discrete state 
$\phi$ and many continuum wave functions $\psi_E^{-(j)}$. The latter
are taken to pre-diagonalize $H$. In other words, it is assumed that
continuum 
wave-functions are  unable to  interact directly with each
other.  They can, however,
interact with each other indirectly via their interactions with the 
discrete state $\phi$ if the following integrals
\begin{equation}
V_{jE} = \left(\psi_E^{-(j)} | H|\phi \right), \label{VjE}
\end{equation}
are not zero. This means that 
$H$ is no longer diagonalized in the combined space spanned by the
basis set $\{ \phi 
, \psi_E^{-(j)} \}$. Eigenfunctions $\Psi_E^{-(j)}$ of $H$ in the
space spanned by $\{ 
\phi , \psi_E^{-(j)} \}$  can be obtained analytically as shown by
Fano\cite{3,4}. By applying the
incoming wave boundary conditions to those eigenfunctions as follows
\begin{equation}
\Psi_E^{-(j)} \rightarrow \sum_{j'} \phi_{j'} (\omega
)\sqrt{\frac{m}{2\pi  
k_{j'}}}
\left( e^{ik_{j'}R} \delta_{j'j} - e^{-ik_{j'}R} S_{j'j} \right)
,           \label{psij_asym}
\end{equation}
S matrix may be obtained\cite{4} as
\begin{equation}
S_{j'j} = \sum_{j''}S_{j'j''}^{0} \left( \delta_{j''j} + 2\pi i
  \frac{V_{j''E} 
  V_{jE}^{\star}}{E-E_{0} - i\pi\sum_k | V_{kE} |^2 } \right) ,
  \label{Smatrix} 
\end{equation}
where $S_{j'j''}^{0}$ is the $S$ matrix of the background
scattering.  
Eq. (\ref{Smatrix}) is different from that of outgoing wave in that 
$i$ is replaced by $-i$. 
The incoming wave instead of outgoing one is employed here as our
interests 
are in the photo-dissociation processes.

As stated in Ref. \cite{3},
$2\pi \sum_k |V_{kE} |^2$ is the spectral width of the resonance peak
and is denoted as $\Gamma$. According to Ref. \cite{5,8},
Eq. (\ref{Smatrix}) 
may be rewritten as
\begin{equation}
S_{j' j} = \sum_{j''j'''} U_{j' j''}^{0} (
e^{-2i\delta_{j''}^{0} } \delta_{j''j'''}
+ic_{j''} c_{j'''} ) 
\widetilde{U_{j''' j}^{0}} . 
\end{equation}
In matrix notation, the above equation becomes
\begin{equation}
S = U^{0} \left( e^{-2i\delta^{0}} + i c \tilde{c} \right)
 \widetilde{U}^{0} 
 \equiv 
U^{0} A \widetilde{U^{0}} , 
\end{equation}
where $c$ is the column vector and its $k$-th element is defined by
\begin{equation}
c_k \equiv e^{-i\delta _k^{0} } \sqrt{ \frac{\Gamma_k}{E-E_{0}
    -i\Gamma/2}} . \label{c_k}
\end{equation}
$\Gamma_k$ is the partial resonance decay width to the
background eigenchannel $k$.

Let us  consider obtaining eigenphase shifts $\delta$ by
the diagonalization of $S$ matrix 
\begin{equation}
S = U e^{-2i\delta} \tilde{U} .\label{S_diag}
\end{equation}
If $A$ = $e^{-2i\delta^{0}}
+ic\tilde{c}$ is diagonalized as
\begin{equation}
A = Ve^{-2i\delta} \tilde{V} , \label{A_def}
\end{equation}
\begin{equation}
S = U^{0} V e^{-2i\delta} \tilde{V} \widetilde{U^{0}}
. \label{S_A_diag} 
\end{equation}
From Eqs. (\ref{S_diag}) and (\ref{S_A_diag}), $U = U^{0}V$. 

Let $x$ denote the eigenvalue  
$e^{-2i\delta}$ of $A$, then the eigenvalue of $A = e^{-2i
\delta^{0}} + 
ic \tilde {c}$ can be obtained by solving the following secular
equation 
\begin{equation}
\left|
\begin{array}{cccc}
e^{-2i\delta_1^{0}} +ic_1 ^2 -x& ic_1 c_2&i c_1 c_3&\cdots\\
ic_2 c_1 & e^{-2i\delta_2^{0}}+ic_2 ^2 -x&ic_2 c_3 &\cdots\\    
ic_3 c_1 &ic_3 c_2 & e^{-2i\delta_3^{0}}+ic_3 ^2 -x&\cdots\\    
\cdot & \cdot & \cdot & \cdots \\
\cdot & \cdot & \cdot & \cdots 
\end{array}
\right| =0  .
\end{equation}
If we divide the first row of the above determinant by $ic_1$, the
second row by $ic_2$ and the $k$-th row by $ic_k$ in general, and 
if we introduce $x_k$ defined by  
\begin{equation}
x_k \equiv \frac{e^{-2i\delta_k^{0}} +ic_k ^2 -x}{ic_k}~~~~~~ ({\rm or}~
x_k -c_k = \frac{e^{-2i\delta_k^{0}}-x}{ic_k } )  ,\label{x_k}
\end{equation}
the secular equation becomes
\begin{equation}
\left| \begin{array}{cccc}
x_1&c_2&c_3&\dots\\
c_1&x_2&c_3&\dots\\
c_1&c_2&x_3&\dots\\
\cdot & \cdot & \cdot & \cdots \\
\cdot & \cdot & \cdot & \cdots 
\end{array} \right| = 0. 
\end{equation}
The above equation for the system of $n$ continua can be transformed  into
 Sardi's form\cite{12}  
\begin{equation}
\prod_{k=1}^{n} (x_k -c_k )+\sum_{k=1}^n \left[ c_k
\frac{\partial}{\partial x_k }\prod_{m=1}^n (x_m - c_m ) \right] =0
. \label{Sardi} 
\end{equation}

By dividing Eq. (\ref{Sardi}) by its first term,
one obtains
\begin{equation}
\sum_{k=1}^n \frac{c_k}{c_k - x_k} = 1 . \label{Sardi2}
\end{equation}
By substitution of Eqs. (\ref{c_k}) and (\ref{x_k}),
Eq. (\ref{Sardi2}) becomes
\begin{equation}
i \sum_{k=1}^n \frac{\Gamma_k}{e^{-2i (\delta - \delta_k^{0} )}-1}
= E -E_{0} -i \frac{\Gamma}{2} . \label{Sardi3}
\end{equation}
Equating the real and imaginary parts, 
Eq. (\ref{Sardi3}) yields two relations: one is Macek's formula
and the other one is $\sum_k \Gamma_k$ = $\Gamma$. This derivation of
Macek's formula is simpler than the original one. 
The Sardi's form is given as a polynomial of order n and more
convenient than Macek's formula in obtaining the solution. In
particular, it shows explicitly that the number of roots of Macek's
formula is the same as the number of continua, $n$. We will use it in
the next section to obtain the solution for the system of one discrete
state and two continua. Before doing that, let us comment on the
general properties of eigenphase shifts obtainable from the Macek's
formula. 

From Macek's
formula, we could draw two properties of eigenphase shifts. 
Firstly, its differentiation with respect to energy
yields 
\begin{equation}
\frac{d\delta_m}{dE} = 2 \left[ \sum_{k=1}^n \frac{\Gamma_k}{\sin ^2
  (\delta_k^{0} - \delta_m )} \right]^{-1} > 0,
  ~~m=1,\ldots,n. \label{deriv_Sardi} 
\end{equation}
The above equation tells us that the first derivatives of eigenphase
shifts with respect to energy near a resonance are positive. 
Secondly,  asymptotic values of eigenphase shifts (at
$|2(E-E_{0})/\Gamma 
| \rightarrow \infty$) are given as $\delta_1^{0}$,
$\delta_2^{0}$, $\delta_3^{0}$, ..., $\delta_n^{0}$ 
up to multiples of $\pi$. 

Let us assume
that $\delta_1^{0} < \delta_2^{0} < \cdots <
\delta_n^{0}$ and that they are all in the range of [0,$\pi$],
for simplicity, which can be always done without loosing generality by
the Hazi's formula (\ref{Hazi}).
Then, the above two properties on eigenphase shifts tell us that
eigenphase shifts vary between two asymptotic values of their
abscissas given by
$[ \delta_1^{0} ,\delta_2^{0} ]$,
$[ \delta_2^{0} ,\delta_3^{0} ]$, $\dots$, 
$[ \delta_{n-1}^{0} ,\delta_n^{0} ]$.
If $\delta_n^{0}$ becomes larger than $\pi$, it is shifted by
$-\pi$ and increases again toward
$\delta_1^{0}$. Since the sum of eigenphase shifts shows the
resonance behavior and increases by $\pi$ around resonance, each
eigenphase shift should similarly show up the resonance behavior
between two consecutive asymptotes of the abscissas $[ \delta_i^{0}
,\delta_{i+1}^{0} ],~~(i=1,\dots , n) $.
When the consecutive eigenphase shifts are the same, let us say
$\delta_i^{0}$ = $\delta_{i+1}^{0}$, then one of two
corresponding eigenphase shifts remains constant while the
other one shows up the resonance behavior between two asymptotes $[
\delta _i^{0}, \delta_{i+2}^{0} ]$.

\section{Eigenphase shifts for the system of one discrete state and two
continua}
\label{sec:eps_one_discrete_two_continua}
  
Let us now consider the system of one discrete state and two continua. 
For this system, Eq. (\ref{Sardi}) becomes
\begin{equation}
[(x_1 - c_1 )(x_2 -c_2 )+c_1 (x_2 -c_2 ) +c_2
(x_1 - c_1 )] =0 .  \label{four_a}
\end{equation} 
Substituting Eq. (\ref{x_k}) into (\ref{four_a}), we obtain
\begin{equation}
x^2 - (e^{-2i \delta_1^0} +ic_1^2 + e^{-2i \delta_2^0} +ic_2^2 )x+
(e^{-2i \delta_1^0} +ic_1^2)(e^{-2i \delta_2^0} +ic_2^2 )
+ {c_1} ^2 {c_2} ^2 =0 . \label{quad_eqn}
\end{equation}

The roots of the quadratic equation (\ref{quad_eqn}) are obtained as
\begin{eqnarray}
x&=&e^{-2i \delta_{\pm}
(E)} \nonumber \\
&=&\frac{e^{-i(\delta_1^{0}+\delta_2^{0})}} 
{E-E_{0}-i \frac{\Gamma}{2}}  
\left( (E-E_{0})\cos \Delta_{12}^{0}+
\frac{\Delta\Gamma}{2}\sin \Delta_{12}^{0}
 \pm i\left\{ \Gamma_1  \Gamma_2 +  \left[ 
(E-E_{0}  )\sin \Delta_{12}^{0} -
\frac{\Delta\Gamma}{2} \cos\Delta_{12}^{0}  
 \right]^2 \right\}^{1/2} \right ) ,
\end{eqnarray}
where  $\Gamma = \Gamma_1 + \Gamma_2$, $\Delta \Gamma = \Gamma_1 -
\Gamma_2 $, and $\Delta_{12}^{0} =
\delta_1^{0}-\delta_2^{0}$.
If we introduce a dimensionless energy $\epsilon_r$ (taken
differently from the usual notation $\epsilon$ in order to avoid the
confusion with another dimensionless energy $\epsilon_a$ which will be
defined 
later) as 
\begin{equation}
\epsilon_r \equiv \frac{E-E_{0}}{\Gamma/2} ,
\end{equation}
the above equation becomes
\begin{eqnarray}
e^{-2 i \bar{\delta}_{\pm} (\epsilon_r )} =
\frac{e^{-i(\delta_1^{0}+\delta_2^{0})}}{\epsilon_r -i}  
\left\{ 
\epsilon_r \cos\Delta_{12}^{0} + \frac{\Delta\Gamma}{\Gamma}
\sin \Delta_{12}^{0} 
\pm  i  
\left[ \frac{4\Gamma_1 \Gamma_2}{\Gamma ^2} +  \left( \epsilon_r
\sin \Delta_{12}^{0} - \frac{\Delta\Gamma}{\Gamma}
\cos \Delta_{12}^{0} \right)^2 \right]^{1/2}
\right\} , \label{solution_2}
\end{eqnarray}
where $\bar{\delta}$ is used instead of $\delta$ in order to emphasize
that the eigenphase shifts are now functions of $\epsilon_r$ instead
of $E$.

Analysis of the eigenphase shifts in Eq. (\ref{solution_2}) is done in
the next subsection by introducing phase shifts due to  the resonance
and avoided crossing interaction. 

\subsection{Phase shifts due to  resonance and avoided crossing
  interaction}

Eq. (\ref{solution_2})  can be decomposed into the product of two terms
of unit modulus\cite{13}:
\begin{eqnarray}
e^{-i \bar{\delta}_r (\epsilon_r ) } &=& \frac{-\sqrt{\epsilon_r ^2
+1}}{\epsilon_r -i} , \\
e^{\pm i \bar{\delta}_a (\epsilon_r )} &=& \frac{-1}{\sqrt{\epsilon_r
^2+1}}   
\left\{ 
\epsilon_r \cos\Delta_{12}^{0} + \frac{\Delta\Gamma}{\Gamma}
\sin \Delta_{12}^{0}  
\mp  i  
\left[ \frac{4\Gamma_1 \Gamma_2}{\Gamma ^2} +  \left( \epsilon_r
\sin \Delta_{12}^{0} - \frac{\Delta\Gamma}{\Gamma}
\cos \Delta_{12}^{0} \right)^2 \right]^{1/2}
\right\} ,  \label{avoiding_term}
\end{eqnarray}
(the notations $\delta_r$ and $\delta_a$ without the bar are also used
elsewhere if it is desired to express them  in terms of $E$ instead of
$\epsilon_r$). 
The phase shift $\bar{\delta}_r$ in the first term is 
the one due to the  resonance as already mentioned in
Eq. (\ref{delta_single}). In single channel scattering, it is the only
term 
besides the background phase 
shift contributing to the eigenphase shifts near a resonance. In  two
channel scattering, there enters another phase shift $\bar{\delta}_a$
defined by the second term due to the indirect coupling of two
continua via quasi-bound state 
near a resonance. 

The functional dependence of $\bar{\delta}_a (\epsilon_r )$ on energy
may be best seen by 
considering its cotangent like $\cot \delta_r = - \epsilon_r $ as in
single channel scattering:
\begin{equation}
\cot \bar{\delta}_a (\epsilon_r) =  \frac{- \left( \epsilon_r
\cos\Delta_{12}^{0} +  
\frac{\Delta\Gamma}{\Gamma} \sin \Delta_{12}^{0} \right) }
{ \left[ \left( \epsilon_r \sin
  \Delta_{12}^{0} - \frac{\Delta\Gamma}{\Gamma} \cos
  \Delta_{12}^{0} \right)^2  +
 \frac{4\Gamma_1 \Gamma_2}{\Gamma ^2} \right]^{1/2}} .
\label{delta_a_primitive} 
\end{equation}
Let us first rewrite Eq. (\ref{delta_a_primitive}) as follows
\begin{equation}
\cot \bar{\delta}_a (\epsilon_r) = - \cot \Delta_{12}^{0} 
\frac{ \frac{\Gamma \sin \Delta_{12}^{0}}{2\sqrt{\Gamma_1
\Gamma_2}} 
\left( \epsilon_r +
\frac{\Delta\Gamma}{\Gamma} \tan \Delta_{12}^{0} \right) }
{ \left\{ \left[ \frac{\Gamma \sin \Delta_{12}^{0}
}{2\sqrt{\Gamma_1 \Gamma_2}} \left( \epsilon_r 
  - \frac{\Delta\Gamma}{\Gamma} \cot
  \Delta_{12}^{0} \right) \right]^2  + 1 \right\}^{1/2}} .
\label{delta_a_primitive_2} 
\end{equation}
The term inside the bracket in the denominator of
Eq. (\ref{delta_a_primitive_2}) can be transformed as follows
\begin{equation}
\frac{\Gamma \sin \Delta_{12}^{0}}{2\sqrt{\Gamma_1 \Gamma_2}}
\left( \epsilon_r - \frac{\Delta \Gamma}{\Gamma} \cot
\Delta_{12}^{0} \right) = \frac{\sin
\Delta_{12}^{0}}{\sqrt{\Gamma_1 \Gamma_2}} \left( E - E_{0}
-\frac{\Delta\Gamma}{2} \cot \Delta_{12}^{0} \right)
. \label{delta_a_denom} 
\end{equation}
This suggests that we can introduce a new energy unit $\Gamma_a$:
\begin{equation}
\Gamma_a \equiv \frac{2\sqrt{\Gamma_1 \Gamma_2}}{ \sin
  \Delta_{12}^{0} } \label{Gamma_a} =
 \frac{\sqrt{\Gamma ^2 - \Delta\Gamma ^2}}{ \sin
\Delta_{12}^{0} } .
\end{equation}
For simplicity, let us also introduce a new parameter $E_a$:
\begin{equation}
E_a \equiv E_{0} + \frac{\Delta \Gamma}{2} \cot
\Delta_{12}^{0} \label{E_a} ,
\end{equation}
(later it will be shown that $E_a$ is the avoided crossing point
energy and  $\Gamma_a$ is
the strength of the
avoided crossing interaction).

With these parameters, Eq. (\ref{delta_a_denom}) becomes
$2(E-E_a )/ \Gamma_a$ which may be considered as a new dimensionless
energy and will be denoted as $\epsilon_a$:
\begin{equation}
\epsilon_a \equiv  \frac{2(E-E_{a})}{\Gamma_{a}} . \label{e_a}
\end{equation}
By substituting Eqs. (\ref{delta_a_denom}) and (\ref{e_a}) into 
(\ref{delta_a_primitive_2}), we obtain
\begin{equation}
\cot \tilde{\delta_a} (\epsilon_a ) = - \cot \Delta_{12}^{0} \cdot
  \frac{\epsilon_a -q_{a} 
  }{\sqrt{\epsilon_a ^2 +1 }} \label{delta_a_tilde} ,
\end{equation}
where $\tilde{\delta}_a (\epsilon_a )$ is used instead of
$\bar{\delta}_a (\epsilon_r )$ in order to emphasize that it is now
a function of $\epsilon_a$ and $q_a$ is the new parameter defined by 
\begin{equation}
q_{a} \equiv \frac{-\Delta\Gamma}{\sqrt{\Gamma ^2 - \Delta \Gamma ^2} \cos
  \Delta_{12}^{0}} . \label{q_a}
\end{equation}
Eq. (\ref{delta_a_tilde}) is a functional of the Beutler-Fano function
defined by
\begin{equation}
f_{{\rm BF}} (\epsilon ,q) \equiv \frac{(\epsilon -q)^2}{1+\epsilon ^2} ,
\end{equation}
and can be rewritten as follows
\begin{equation}
\cot \tilde{\delta_a} (\epsilon_a ) = 
\left\{ \begin{array}{ll}
\cot \Delta_{12}^{0} \sqrt{f_{{\rm BF}} (\epsilon_a , q_{a} )}  &
  {\rm when}~\epsilon_a 
< q_a \\ 
- \cot \Delta_{12}^{0} \sqrt{f_{{\rm BF}}  (\epsilon_a , q_{a} )} 
& {\rm when}~\epsilon_a \ge q_a 
\end{array} \right. , \label{delta_a_bf}
\end{equation}
Eq. (\ref{delta_a_tilde}) or (\ref{delta_a_bf}) tells us that the
phase shift $\delta _a$ due to the avoided crossing interaction takes
its simplest form when it is parameterized in terms of $\epsilon_a$
and $q_a$.
Since we want $\Gamma_a$ to be positive, it will be assumed that
$\sin\Delta_{12}^{0}\ge 0$, which can always be achieved by the
appropriate choice of 1 and 2 for $\delta_1^{0}$ and
$\delta_2^{0}$. With this convention in mind,
$\Gamma_a$ will be written as
$2 \sqrt{\Gamma_1 \Gamma_2 }/ |\sin \Delta_{12} |$
hereafter.  

With the above phase shifts, eigenphase shifts are obtained as 
\begin{equation}
2\delta_{\pm} (E) = \delta_1^{0} + \delta_2^{0} + \delta_r (E)
\pm \delta_a (E) , \label{delta} 
\end{equation}
(for the choice of the sign, see \cite{13}).
It is interesting to note that 
phase shifts of different origins are added up linearly to eigenphase
shifts. 

Using Eq. (\ref{Hazi}), the above equation may be expressed in terms
of the eigenphase sum $\delta_{\Sigma} (E)$
\begin{equation}
2\delta_{\pm} (E) = \delta_{\Sigma} (E) \pm \delta_a (E)
. \label{delta2}
\end{equation}
The eigenphase sum $\delta_{\Sigma} (E)$ consists of the
background eigenphase sum $\delta_{\Sigma}^{0}$ and the phase
shift $\delta_r (E)$ due to the resonance. The phase shifts due to the
resonance contribute positively from 
zero to $\pi$ as energy varies around a resonance. The phase shift is
positive since particles are
attracted and bound temporarily near  a resonance. 
The $\pm$ sign in front of $\delta _a (E)$ in Eq. (\ref{delta}) tells
us that the phase shift due to the avoided  
crossing interaction in one eigenchannel increases at the expense
of the phase shift of the other eigenchannel and vice versa. If it is
increased 
in one  eigenchannel due to attraction, it decreases in the other
one  due to repulsion. They are exactly canceled out and do not
contribute to the  eigenphase sum.
The study of $\delta_a$ as a function of $\epsilon_a$ shows that
$\delta_a$ has an extremum of $\cot^{-1}  (\sqrt{q_a^2 +1} \cot
\Delta_{12}^{0} )$ at $\epsilon_a = - 1/q_a$ (a minimum when $q_a
>0$, a maximum when $q_a <0$).  
From its definition, the limiting behavior of $\tilde{\delta}_a$ at
off-resonance is obtained as 
\begin{equation}
\tilde{\delta}_{a} (\epsilon _a ) \rightarrow 
\left\{ \begin{array}{ll}
 \Delta_{12}^{0}& {\rm when}~\epsilon_a \rightarrow -\infty ,\\
 \pi - \Delta_{12}^{0}& {\rm when}~\epsilon_a \rightarrow \infty ,
\end{array} \right.  \label{delta_a_limits}
\end{equation}
[note that $0 \le \Delta_{12}^{0} \le \pi /2$ according to the 
convention mentioned below Eq. (\ref{delta_a_bf})].
$\tilde{\delta}_a$ increases around resonance 
by $\pi - 2\Delta_{12}^{0}$. Thus the phase shift due to avoided
crossing interaction varies most around resonance when
$\Delta_{12}^{0}$ = 0 and least
when $\Delta_{12}^{0}$ = $\pi /2$. 
It passes the middle, $\pi /2$, of two asymptotes of the abscissas in
Eq. (\ref{delta_a_limits}) when $\epsilon_a$ = $q_{a}$. 
Since the avoided crossing point energy $E_a$
corresponds to zero of $\epsilon_a$ [see the discussion below Eq. 
(\ref{trig_theta_a})], 
$q_{a}$ tells how the avoided
crossing point energy is apart from the middle of two asymptotes of
the abscissas. Such
an apartness is a measure of the asymmetry of the curves of
$\tilde{\delta}_a (\epsilon_a )$. 
Fig. \ref{fig:delta} 
shows the variations
in the behaviors of $\delta_a$ and $\delta_{\pm}$
as $q_a$ varies.

So far, we obtained eigenvalues of $S$ matrix for two open channels
near a resonance and decomposed them into contributions from the
background, resonance, and avoided crossing interactions. 
Let us now consider obtaining the eigenvectors of $S$ matrix which are
usually called  eigenchannels. $S$ matrix is diagonalized in two
steps. It is first transformed to $A$ by $U^{0}$ which diagonalizes 
$S^{0}$ matrix, i.e., $S$ = 
$U^{0} A \widetilde{U^{0}}$ 
($S^{0} = U^{0}
e^{-2i\delta^{0}} \widetilde{U^{0}}$). $A$ is then
diagonalized as in Eq. (\ref{A_def}) by $V$ matrix which is composed
of two 
eigenvectors $v_{+}$ and $v_{-}$ corresponding to $\delta_{+}$ and
$\delta_{-}$ as
\begin{equation}
V = \left( \begin{array}{cc} v_{+} & v_{-}\end{array} \right) .
\end{equation}
After a lengthy derivation, eigenvectors  are obtained as 
\begin{equation}
v_{+} = \left[ \begin{array}{c}
\cos (\frac{\theta_{a}}{2}) \\
 \sin (\frac{\theta_{a}}{2})  
\end{array} \right] ,  ~~
v_{-} = \left[ \begin{array}{c}
- \sin (\frac{\theta_{a}}{2}) \\
\cos (\frac{\theta_{a}}{2}) 
\end{array} \right] , \label{eigenvector} 
\end{equation}
with $\theta_{a}$  defined by 
\begin{equation}
\cos\theta_{a} \equiv -\frac{\epsilon_a}{\sqrt{1+\epsilon_a
    ^2}} , ~~
\sin\theta_{a} \equiv \frac{1}{\sqrt{1+\epsilon_a
    ^2}} .  \label{trig_theta_a}
\end{equation}
It is interesting to note that eigenvectors are independent of
$q_{a}$. They depend only on $\epsilon_a$. As
$\epsilon_a$ varies from $-\infty$ through zero to $\infty$,
$\theta_{a}$ varies from zero through $\pi/2$ to $\pi$ and
$v_{+}$ varies from $\left( \begin{array}{cc}1\\ 0 \end{array} \right)$
through 
$\frac{1}{\sqrt{2}} \left( \begin{array}{cc}1\\ 1 \end{array} \right)$
to $\left( \begin{array}{cc}0\\ 1 \end{array} \right)$ .
Thus, at
$\epsilon_a$ = 0 or at $E$ = $E_{0}$ + $\frac{\Delta\Gamma}{2}\cot
\Delta_{12}^{0}$, two eigenphase shifts are avoided most. For
this reason $\epsilon_a$ = 0 is considered as the avoided crossing
point 
energy. (Another way to see that $\epsilon_a = 0$ is the avoided
crossing point energy might be to remove the cause of the avoidance of
two eigenphase shift curves and to let them cross through each other
and to see whether the crossing point is $\epsilon_a = 0$. This was
done and $\epsilon_a =0$ is confirmed to be the crossing point.)

Let us parameterize $U^{0}$ matrix  as
\begin{equation}
 U^{0} = \left( \begin{array}{cc} 
\cos\theta^{0} & - \sin  \theta^{0} \\
\sin  \theta^{0}& \cos\theta^{0}  \end{array} \right) ,
\label{u_0} 
\end{equation}
where $\theta^{0}$ may be considered as the  angle by which the
frame for 
background eigenchannels is rotated from the frame for background
asymptotic channels. By calculating the multiplication of $U =
U^{0}V$ and by noting that $U$ matrix is composed of eigenvectors
$u_{+}$ and $u_{-}$ of $S$ matrix as 
\begin{equation}
U = \left( \begin{array}{cc}u_{+}& u_{-} \end{array} \right) ,
\end{equation}
the eigenvectors of $S$ matrix are obtained as
\begin{equation}
u_{+} = \left[ \begin{array}{c}
\cos (\frac{\theta_a'}{2}) \\
\sin (\frac{\theta_a'}{2})  
\end{array} \right] ,  ~~
u_{-} = \left[ \begin{array}{c}
- \sin (\frac{\theta_a'}{2}) \\
\cos (\frac{\theta_a'}{2}) 
\end{array} \right] ,  \label{u_eigenvector}
\end{equation}
where $\theta_a'$ is defined by 
\begin{equation}
\theta_a' \equiv \theta_a + 2\theta^{0}  .  \label{theta_ap}
\end{equation}

\subsection{$S$ matrix in terms of new parameters}

Let us now express $S$ matrix in terms of new parameters introduced in
the previous section.
Reassembling by using its eigenvalues 
$e^{-2i\delta_{+}}$ and $e^{-2i\delta_{-}}$
($\delta_{\pm}$ = $\delta_{\Sigma}$ $\pm \delta_a$) and eigenvectors
Eq. (\ref{eigenvector}), $A$ becomes as follows
\begin{eqnarray}
 A &=&
\left( \begin{array}{cc} v_{+} & v_{-} \end{array} \right) 
\left( \begin{array}{cc} e^{-2i\delta_{+}} & 0\\
0&e^{-2i\delta_{-}}\end{array} 
\right) 
\left( \begin{array}{c}  \widetilde{v_{+}} \\ \widetilde{v_{-}}
\end{array} \right) \nonumber \\ 
&=& e^{-i \delta _{\Sigma} } \left( \begin{array}{cc}
\cos \delta_a -i \sin\delta_a \cos \theta_a & - i\sin \delta_a
\sin\theta_a \\ 
- i\sin \delta_a \sin\theta_a &
\cos \delta_a +i \sin\delta_a \cos \theta_a 
\end{array} \right)  .
\end{eqnarray}

From the relation $S = U^{0} A
\tilde{U^{0}}$ with
$U^{0}$ matrix defined in (\ref{u_0}), $S$ matrix is obtained as
\begin{equation}
 S = e^{-i \delta _{\Sigma} } \left( \begin{array}{cc}
\cos \delta_a -i \sin\delta_a \cos \theta_a'  & - i\sin \delta_a
\sin\theta_a'  \\ 
- i\sin \delta_a \sin\theta_a' &
\cos \delta_a +i \sin\delta_a \cos \theta_a'
\end{array} \right)  ,
\end{equation}
where $\theta_a'$ is defined in Eq. (\ref{theta_ap}).
Using Pauli matrices, it can be transformed into an invariant form:
\begin{eqnarray}
 S &=& e^{-i \delta _{\Sigma} } 
\left\{ \cos\delta_a  -i \sin\delta_a [ \sigma_z 
\cos \theta_a' + \sigma_x \sin \theta_a'
] \right\} \\
&=& e^{-i \delta _{\Sigma} } \left( \cos\delta_a -i \sin\delta_a 
\vec{\sigma} \cdot \hat{n}_{\theta_a'}  \right) , \label{smat_pauli}
\end{eqnarray}
where the unit vector $\hat{n}_{\theta_a'}$ is defined as 
\begin{equation}
 \hat{n}_{\theta_a'} \equiv \hat{z} \cos \theta_a' + \hat{x}
\sin\theta_a'  .   \label{n_thetap}
\end{equation} 
Since
\begin{equation}
\cos\delta_a -i \sin\delta_a 
\vec{\sigma} \cdot \hat{n}_{\theta_a'} = e^{-i\delta_a \vec{\sigma}
  \cdot \hat{n}_{\theta_a'}} , 
\end{equation}
it is simplified into the form of
$e^{iH}$ ($H$ 
hermitian):
\begin{equation}
S = e^{-i\left(  \delta_{\Sigma} +\delta_a 
\vec{\sigma} \cdot \hat{n}_{\theta_a'}  \right) } .
\end{equation}

Since $\left( \vec{\sigma} \cdot \hat{n}_{\theta_a'} \right)^2$ = 1,
$\vec{\sigma} \cdot \hat{n}_{\theta_a'}$ has two eigenvalues
$\pm1$. $u_{+}$ and $u_{-}$ in Eq. (\ref{u_eigenvector}) are their
corresponding eigenvectors. In this case, it is well known\cite{14}
that the vectors 
$\pm \hat{n}_{\theta_a'}$ equal the expectation values of the Pauli spin
operator, $( u_{\pm} | \vec{\sigma} | u_{\pm} )$, and are called the
polarization vectors that correspond to 
the eigenstates $u_{\pm}$, respectively. 
We can obtain the change in the direction of
the polarization axis as a function of energy by differentiating
Eq. (\ref{n_thetap}) with respect to $\epsilon_a$:
\begin{equation}
\frac {d\hat{n}_{\theta_a'}}{d\epsilon_a } = \hat{y} \times
\hat{n}_{\theta_a'} \frac{d\theta_a'}{d\epsilon_a}  .
\end{equation}
Since $\tan \theta_a = -1/\epsilon_a$, 
\begin{equation}
\frac{d\theta_a' }{d\epsilon_a} = \frac{d\theta_a }{d\epsilon_a} =
\frac{1}{1+\epsilon_a^2} 
\label{dthetade} 
\end{equation}
The above equations also appear in the spin model for the adiabatic
analysis of collisions\cite{15}.

\section{Times delayed by collision near a resonance}
\label{sec:etd}

\subsection{Time delay matrix and eigentime delays}

For a single channel system, times delayed by collision, ${\tau}$, are
obtained by the first derivatives of 
phase shifts:
\begin{equation}
{\tau} (E)= 2 \hbar \frac{d\delta (E)}{dE}.
\end{equation}
For a multichannel system, we may consider the time delay matrix $Q$
defined 
by\cite{9,10} 
\begin{equation}
Q  = i\hbar
 \left( S^{\dag} \frac{dS}{dE}\right) , \label{TDdef}
\end{equation}
The unitarity of $S$ matrix ensures that the time delay matrix $Q$ is
hermitian. Let us now consider the time delay matrix for the system of
one discrete state and two continua. Using Eq.  (\ref{smat_pauli}) for
$S$ matrix 
expressed in terms 
of the Pauli matrices,  the first derivative of $S$ matrix with
respect to energy is calculated as
\begin{eqnarray}
 \frac{dS}{dE} = -i\frac{d\delta_{\Sigma}}{dE} S 
+ e^{-i\delta_{\Sigma}} 
\left( -\sin\delta_a -i\cos\delta_a
\vec{\sigma} \cdot \hat{n}_{\theta_a'} \right) \frac{d\delta_a}{dE}
+ e^{-i\delta_{\Sigma}} 
\left( -i \sin\delta_a \vec{\sigma} \cdot
  \frac{d\hat{n}_{\theta_a'}}{dE} \right) .
\end{eqnarray}
Now the time delay matrix $Q$ is obtained as
\begin{eqnarray}
i\hbar S^{\dag} \frac{dS}{dE} = \hbar \frac{d\delta_{\Sigma}}{dE} +
\vec{\sigma} \cdot \hat{n}_{\theta_a'} \hbar \frac{d\delta_a}{dE}
+ \sin \delta_a \vec{\sigma} \cdot \left[ 
\hat{y} \times \hat{n}_{\theta_a'} \cos \delta_a
- \hat{y}  \sin\delta_a \right] \left( \hbar
\frac{d\theta_a'}{dE} \right)  .
\end{eqnarray}
Since $\hat{n}_{\theta_a'} \times \hat{y}$ and $\hat{y}$ are
perpendicular to $\hat{n}_{\theta_a'}$, their linear combination is
also perpendicular to $\hat{n}_{\theta_a'}$. Let us introduce a
vector
\begin{equation}
\hat{n}_{\theta_a'}^{\perp} \equiv \hat{y} \times \hat{n}_{\theta_a'}
 \cos \delta_a  - \hat{y} \sin\delta_a  ,
\end{equation}
perpendicular to $\hat{n}_{\theta_a'}$ and making an angle $\delta_a$
with $\hat{n}_{\theta_a'} \times \hat{y}$
and also define the following quantities
\begin{eqnarray}
\tau_r &\equiv& 2\hbar  \frac{d\delta_r }{dE} , \nonumber\\
\tau_a &\equiv& 2\hbar  \frac{d\delta_a}{dE} , \nonumber\\
\tau_f &\equiv& 2\hbar \sin\delta_a \frac{d\theta_a'}{dE} . \label{tau_f}
\end{eqnarray}
The last three quantities correspond to the time delay due to the
resonance,  
the avoided crossing interaction, and
the change of the direction of
the eigenvectors $\hat{n}_{\theta_a'}$ of $S$ matrix, or usually
termed as the frame change, as a function of
energy, 
respectively.  
Then, the time delay matrix $Q$ may be written as
\begin{equation}
Q = \frac{1}{2} (\tau_r + 
\vec{\sigma}\cdot \hat{n}_{\theta_a'} \tau_a 
+ \vec{\sigma}\cdot \hat{n}_{\theta_a'}^{\perp} \tau_f )
. \label{Q_pauli} 
\end{equation}
Since observable properties of a particle with spin $\frac{1}{2}$ can
be represented as a function of its spin polarization $\vec{P}$ and
the states of all the two-level (two-continua here) systems can be
mapped on the states of orientation of a particle with spin
$\frac{1}{2}$, $Q$ may be rewritten in terms of the polarization vectors
as 
\begin{equation}
Q = \frac{1}{2} \tau_r \left( 1 + \vec{P_a} \cdot \vec{\sigma} +
\vec{P_f} \cdot \vec{\sigma} \right) ,
\end{equation}
where 
polarization vectors are defined as
\begin{equation}
\vec{P_a} \equiv \frac{\tau_a}{\tau_r} \hat{n}_{\theta_a '} , ~~~ 
\vec{P_f} \equiv \frac{\tau_f}{\tau_r} \hat{n}_{\theta_a '}^{\perp}
. \label{Q_pauli_pol}
\end{equation}
The degrees of polarization $| \vec{P_a} |$ and $| \vec{P_f} |$ will
be denoted simply as $P_a$ and $P_f$. They are positive by definition
(the term polarization may not be adequate for the time delay
matrix. We will return to this point later). 
Let us introduce a new unit vector 
\begin{equation}
\hat{n}_t = \hat{n}_{\theta_a'}
\frac{\tau_a}{\sqrt{\tau_a ^2 + \tau_f ^2}} 
+ \hat{n}_{\theta_a'}^{\perp} \frac{\tau_f}{\sqrt{\tau_a ^2 + \tau_f
    ^2}} .
\end{equation}
Then
Eqs. (\ref{Q_pauli}) and (\ref{Q_pauli_pol}) become
\begin{equation}
Q = \frac{1}{2} \left( \tau_r + 
\vec{\sigma} \cdot \hat{n}_{t} \sqrt{\tau_a ^2 + \tau_f ^2 }
\right) 
= \frac{1}{2} \tau_r \left( 1 + \vec{P_t} \cdot \vec{\sigma} 
\right) ,
\end{equation} 
where
\begin{equation}
\vec{P_t} \equiv \sqrt{P_a ^2 + P_f ^2 } \hat{n}_t .
\end{equation}
Now eigenvalues of the time delay matrix $Q$ or eigentime delays,
which will be denoted as $\tau_{\pm}$, are easily obtained as 
\begin{equation}
\tau_{\pm} = \frac{1}{2} \left( \tau_r \pm \sqrt{\tau_a ^2 + \tau_f
^2} \right)  
= \frac{1}{2} \tau_r \left( 1 \pm P_t \right) .  \label{eigen_time_delays}
\end{equation}
Notice that $\tau_a$ and $\tau_f$ are added up incoherently to
eigentime delays. 
Also because of
${\rm tr} (\sigma_i ) = 0$ ($i$=x,y,z), 
\begin{equation}
{\rm tr} Q =  \tau_{+} + \tau_{-} = \tau_r ,
\end{equation}
that is, the eigentime delay sum has the same
form as that of the 
eigentime delay in the single channel scattering\cite{16}.

\subsection{Explicit formulas of time delays}
\label{sec:td_formulas}

The  time delay due to the resonance, the first term of
Eq. (\ref{eigen_time_delays}), is obtained by differentiating $\tan
\bar{\delta _r } = -1/ \epsilon_r$ with respect to $\epsilon_r$ as
\begin{equation}
\bar{\tau}_r (\epsilon_r ) \equiv 2\hbar \frac{d\bar{\delta}_r (\epsilon_r
  )}{d\epsilon_r}  = \frac{2\hbar}{1+\epsilon_r ^2}.
\end{equation}
It takes the Lorentzian form and is thus always positive or delayed as
it is kept temporarily  
by the quasi-bound state.

The form of the time delay due to avoidance, the second term of
Eq. (\ref{eigen_time_delays}), is rather complex: 
\begin{equation}
\tau _a  (E) =  2\hbar \frac{d\delta_a}{dE} =  \frac{- 2\hbar \left[ 
 \frac{1}{2}\Delta\Gamma (E-E_{0})
\sin\Delta_{12}^{0}
-\left(\frac{\Gamma}{2}
\right)^2 \cos \Delta_{12}^{0} \right]}{\left[(E-E_{0})^2 +
\left(\frac{\Gamma}{2}\right)^2 \right] \left\{ \Gamma_1 \Gamma_2 + 
\left[(E-E_{0} )\sin\Delta_{12}^{0} -\frac{1}{2} \Delta \Gamma
  \cos \Delta_{12}^{0} \right]^2 \right\} ^{1/2} }
. \label{tau_a_E} 
\end{equation}
By changing the independent variable from $E$ to $\epsilon_r$,
considering the reduced time delays $\bar{\tau}_a$ 
(=$2\hbar \frac{d\delta_{a}}{d\epsilon_r}$),
and introducing the new
parameters $r^2$  and $q_{\tau}$ defined by
\begin{equation}
r^2 \equiv \frac{\Gamma ^2 - \Delta\Gamma ^2 }{\Delta \Gamma ^2}
,\label{r_2} 
\end{equation}
\begin{equation}
q_{\tau} \equiv \frac{\Gamma}{\Delta\Gamma} \cot\Delta_{12}^{0} ,
\label{qtau} 
\end{equation}
Eq. (\ref{tau_a_E}) is simplified as
\begin{equation}
\bar{\tau}_a  (\epsilon_r ) = 
- \bar{\tau}_r (\epsilon_r ) \frac{\epsilon_r - q_{\tau}}
{\sqrt{( \epsilon_r - q_{\tau} )^2 + r^2 (1+\epsilon_r ^2 )}}
. \label{tau_a}  
\end{equation}
The factor multiplying $\bar{\tau _r} (\epsilon_r )$ in the above
equation can be transformed to the
functional of the Beutler-Fano 
formula [the Lorentzian shape of $\bar{\tau _r} (\epsilon _r )$ may
be regarded  
as a Beutler-Fano shape]
\begin{eqnarray}
\bar{\tau}_a  (\epsilon_r ) &=& 
\left\{ \begin{array} {ll}
 \bar{\tau}_r (\epsilon_r )
\frac{1}{\sqrt{1+r^2 \frac{1+\epsilon_r^2}{(\epsilon_r - q_{\tau}
      )^2}}},&{\rm when}~\epsilon_r \le q_{\tau},\\
-  \bar{\tau}_r (\epsilon_r )
\frac{1}{\sqrt{1+r^2 \frac{1+\epsilon_r^2}{(\epsilon_r - q_{\tau}
      )^2}}},&{\rm when}~\epsilon_r > q_{\tau} ,
\end{array} \right.  \nonumber \\
&=&
\left\{ \begin{array}{ll}
 \bar{\tau}_r (\epsilon_r )
\sqrt{\frac{f_{{\rm BF}} (\epsilon_r ,q_{\tau})}{f_{{\rm BF}} (\epsilon_r
,q_{\tau} ) +r^2 }},
&{\rm when}~\epsilon_r \le q_{\tau},\\
- \bar{\tau}_r (\epsilon_r )
\sqrt{\frac{f_{{\rm BF}} (\epsilon_r ,q_{\tau})}{f_{{\rm BF}} (\epsilon_r
,q_{\tau} ) +r^2 }},
&{\rm when}~\epsilon_r > q_{\tau} ,
\end{array} \right.  \nonumber \\
&\equiv& \bar{\tau}_r (\epsilon_r ) g_a(\epsilon_r ) . \label{tau_a_bf}
\end{eqnarray}
Interestingly, $\bar{\tau}_a$ takes the Beutler-Fano formula in the
energy scale of $\epsilon_r$ instead of $\epsilon_a$ in contrast to
the case of $\tilde{\delta}_a$, though 
the former is obtained as the derivative of the latter.  
Notice that the absolute value of $g_a$ is the polarization $P_a$
considered in the previous section:
\begin{equation}
P_a = \sqrt{\frac{f_{\rm BF} (\epsilon_r ,q_{\tau})}{f_{{\rm BF}}
(\epsilon_r ,q_{\tau} ) +r^2 }} . \label{P_a}
\end{equation}
$g_a$ changes its sign at $\epsilon_r$
=  $q_{\tau}$, so does the polarization vector $\vec{P}_a$ at the same
energy. The study on $P_a$ as a function of $\epsilon_r$ shows
that $P_a$ has a maximum of $\sqrt{(1+q_{\tau}^2 )/(1+q_{\tau}^2 +r^2
)}$ at $\epsilon_r = -1/ q_{\tau}$ and a minimum of 0 at $\epsilon_r =
q_{\tau}$ (which corresponds to $\epsilon_a = -1/q_a$ where $\delta_a$
has an extremum). 
Fig. \ref{fig:g_td_a} 
shows the variations
in the behaviors of $g_a (\epsilon_r )$ and $\bar{\tau}_a (\epsilon_r
)$ as functions of $\epsilon_r$ at three different profile indices
$q_{\tau}$  
with $\Delta_{12}^{0}$ fixed (the value of $q_{\tau}$ is
restricted by $| q_{\tau} / \cot
\Delta_{12}^{0} | \ge 1$).

Recalling that $f_{\rm BF} (\epsilon_r , q_{\tau} ) \ge 0$,
Eq. (\ref{P_a}) tells us that  
the magnitude of $P_a (\epsilon_r)$ is smaller than or equal to unity and
has the effect of making the absolute magnitude of the time delay
$\bar{\tau}_a $  due to an avoidance  smaller than or equal to that of
the time 
delay $\bar{\tau}_r $ due to a resonance:
\begin{equation}
\bar{\tau}_r (\epsilon_r )  \ge \left| \bar{\tau}_a (\epsilon_r )
\right| ,  \label{tau_inequality}
\end{equation}
which ensures that 
$\bar{\tau}_{r} (\epsilon_r ) \pm \bar{\tau}_a$ are positive.

The above inequality may be obtained without knowing the explicit
functional dependences of $\tau_r$ and $\tau_a$ as functions of
energy.  Notice that 
\begin{equation}
\tau_r \pm \tau_a = 2 \hbar \frac{d ( \delta_r \pm \delta_a )}{dE} =
2\hbar \frac{d \delta_{\pm} }{dE} .
\end{equation}
Because of the inequality (\ref{deriv_Sardi}), the above equation
should be larger than or equal to zero, which proves
(\ref{tau_inequality}).  This inequality is the manifestation  of the
physical fact that eigenphase shifts are increasing functions of 
energy in the neighborhood of a resonance as colliding particles are
kept bound temporarily around a resonance (attractive forces make
positive phase shifts). 

The above inequality restricts the magnitude of $P_a$ to $0\le P_a \le
1$. The magnitude of $P_a$ causes the difference between two eigentimes
delayed.  Eq. (\ref{P_a}) tells us that the
degree of difference in two eigentime delays is governed  by $r^2$. 
Let us consider $r^2$ as a function of $x = \Gamma_1 / \Gamma$ ($0 \le
x \le 
1$).  Then since $r^2$ is symmetric with respect to $x$ =
$\frac{1}{2}$ and its 
values are zero at $x=0$ and 1, infinity at $x=\frac{1}{2}$, the
magnitudes of $P_a$  are 1 at $x=0$ and  0 at $x=\frac{1}{2}$. Since
avoided crossing interaction is strongest at $x=\frac{1}{2}$, the
difference in two eigentimes delayed due to an avoided crossing
interaction completely disappears when the
avoided crossing interaction is strongest.

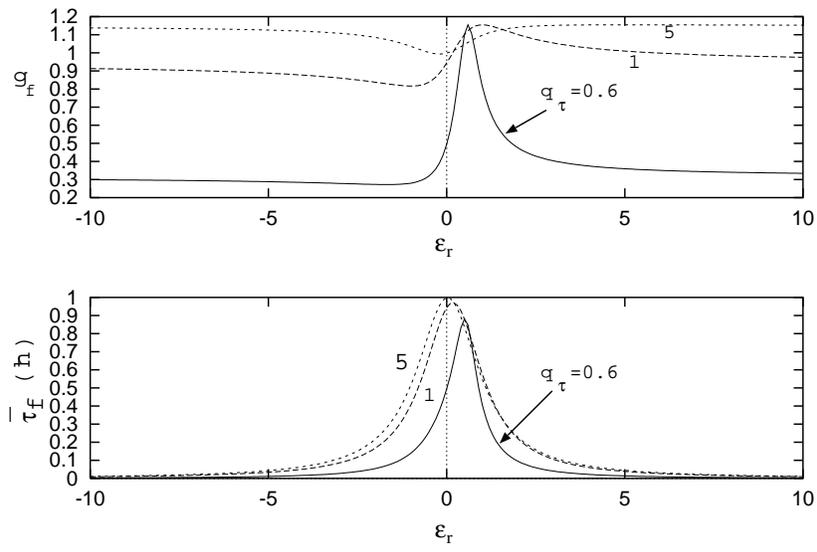
\begin{figure}
\epsfig{file=tau_f.eps,width=11cm}
\caption{$g_f (\epsilon_r )$ and $\bar{\tau}_f (\epsilon_r )$ 
vs. $\epsilon_r$ are plotted for three different profile indices 
$q_{\tau}$ = 0.6, 1,
and 5 with $\Delta_{12}^{0}$ = $\pi /6$.~~~~~~~~~~~~~~~~~~~~~~~~~~~~
}
\label{fig:td_sf}
\end{figure}

Let us now consider obtaining the explicit formula of the times
delayed  due to a frame change.
By substituting Eq. (\ref{dthetade}) and 
\begin{equation}
\sin \delta_a = \frac{\Gamma_a}{\Gamma} \sin \Delta_{12}^{0}
\sqrt{\frac{\epsilon_a^2 +1}{\epsilon_r^2 +1}},
\end{equation}
into Eq. (\ref{tau_f}), the time delay $\tau_f$ due to the frame
change is obtained as 
\begin{equation}
\tau_f = \tau_r \sin \Delta_{12}^{0} \sqrt{\frac{\epsilon_r^2
+1}{\epsilon_a^2 +1}} \equiv 
\tau_r g_f
. \label{tau_f_2} 
\end{equation}
The absolute value of $g_f$ is the same as $P_f$:
\begin{equation}
P_f = | \sin \Delta_{12}^{0} |
\sqrt{\frac{\epsilon_r^2+1}{\epsilon_a^2 +1}} , \label{P_f}
\end{equation}
[actually $g_f$ equals $P_f$ in the convention mentioned below
Eq. (\ref{Gamma_a})].
Fig. \ref{fig:td_sf} 
shows the variations
in the behaviors of $P_f (\epsilon_r )$ and $\bar{\tau}_f (\epsilon_r
)$ as functions of energy at three different profile indices
$q_{\tau}$  with $\Delta_{12}^{0}$ fixed.

As in the  case of $\tau_a$,
$P_f$ is smaller than or equal to unity.
This can be proved by examining the behavior of the graph of $P_f$ as
a function of 
$\epsilon_r$. $P_f$  can be easily transformed as a
function of $\epsilon_r$ by substituting 
\begin{equation}
\epsilon_a = \frac{\Gamma}{\Gamma_a} \left( \epsilon_r - \frac{\Delta
\Gamma}{\Gamma} \cot \Delta_{12}^{0} \right) , \label{epsilon_a2}
\end{equation}
for $\epsilon_a$ into Eq. (\ref{tau_f_2}). It is then differentiated
with respect to energy to yield the roots of  the first derivative of
$\bar{\tau}_f (\epsilon_r )$ at $\epsilon_r$ =
$-1/ q_{\tau}$ and $q_{\tau}$.  From this, it is found that 
$P_f$ has the minimum of $\sin \Delta_{12}^{0} \sqrt{1 - (
\Delta\Gamma / \Gamma )^2}$ at $-1/q_{\tau}$ and the maximum of unity at
$q_{\tau}$. Its values
become $\sqrt{1 - ( \Delta\Gamma / \Gamma )^2}$ at $\epsilon_r
\rightarrow \pm \infty$. This proves $0 \le P_f \le 1$. 

Now let us consider the magnitude of the total polarization vector
$P_t = \sqrt{P_a^2 + P_f^2}$. In order to obtain this, let us further
transform Eq. (\ref{epsilon_a2}) using the definition
(\ref{Gamma_a}) of $\Gamma_a$ and $\Delta\Gamma / \Gamma = \cot
\Delta_{12}^{0} / q_{\tau}$ as follows
\begin{equation}
\epsilon_a = \frac{| \sin\Delta_{12}^{0} |}{\sqrt{1 - \left(
\frac{\cot\Delta_{12}^{0}}{q_{\tau}} \right)^2 }} \left(
\epsilon_r - \frac{\cot ^2 \Delta_{12}^{0}}{q_{\tau}} \right) .
\label{epsilon_a} 
\end{equation}
Also after some manipulation, we
obtain the following relation:
\begin{equation}
(\epsilon_r - q_{\tau} )^2 + r^2 (\epsilon_r^2 +1 ) = \left(
\frac{q_{\tau}}{\cot \Delta_{12}^{0}} \right)^2 \left[ \left(
\epsilon_r - \frac{\cot ^2 \Delta_{12}^{0}}{q_{\tau}} \right)^2 +
\frac{1- \left( \frac{\cot\Delta_{12}^{0}}{q_{\tau}} \right)^2
}{\sin ^2 \Delta_{12}^{0}} \right] .  \label{denom_fb}
\end{equation} 
Using Eq. (\ref{epsilon_a}) to calculate $\epsilon_a^2 +1$ and
comparing the latter with the right hand side of Eq. (\ref{denom_fb}),
we obtain  
\begin{equation}
\epsilon_a^2 + 1 = 
\frac{\sin ^2 \Delta_{12}^{0}}{r^2} \left[ \left( \epsilon_r - 
q_{\tau} \right) ^2 + r^2 \left( \epsilon_r ^2 +1  \right) \right]   ,
\label{e2_p_1}
\end{equation}
where $r^2 = ( q_{\tau} / \cot \Delta_{12}^{0} )^2 -1$ is used.
Substituting Eq. (\ref{e2_p_1}) into Eq. (\ref{P_f}), $P_f$ can be put
into a functional of Beutler-Fano formula:
\begin{equation}
P_f = \frac{r}{\sqrt{f_{\rm BF} (\epsilon_r ,q_{\tau}) + r^2}} ,
\label{P_f2} 
\end{equation}
From Eqs. (\ref{P_f2}) and (\ref{P_a}), we obtain the surprising
result: 
\begin{equation}
P_a^2 + P_f^2 = 1 .
\end{equation}
This means that eigentime delays for the system of one discrete state
and two  continua are zero and $\tau_r (\epsilon_r )$. 
Though time delays due to the avoided crossing interaction and frame
change are asymmetric with respect to the resonance energy and
therefore the energy of the longest lifetime does not match with the
resonance energy, the energy of the longest overall eigentimes delayed
is exactly matched with the resonance energy.

\section{Results and Discussion}
\label{sec:results}

Let us summarize the results. In this paper, we considered eigenphase
shifts and eigentime delays near a resonance for a system of one
discrete state and two continua using Fano's configuration interaction
theory. The eigenphase shifts are obtained as 
\[
2\delta_{\pm} (E) = \sum_i \delta_i^{0} + \delta_r (E) \pm
\delta_a (E) ,
\]
where $\delta_r (E)$ is the phase shift due to the modification of
the scattering wave by the quasi-bound state and given by
$- \arctan (1/\epsilon_r )$ and $\delta_a (E)$ is
the one due to the modification of the scattering wave by the other
wave  through the indirect interaction via the quasi-bound state and
given as a functional of the Beutler-Fano formula:
\[
\cot \delta_a (E) =   - \cot \Delta_{12}^{0}  
\frac{\epsilon_a -q_{a} }{(1+\epsilon_a^2 )^{1/2}}  .
\]
In the above formula, energy is expressed in the unit of
$\Gamma_a  = 2 \sqrt{\Gamma_1 \Gamma_2}/ | \sin\Delta_{12}^{0} |$ 
and shifted from the resonance energy $E_{0}$ by
$\frac{\Delta\Gamma}{2}\cot\Delta_{12}^{0}$. The shape of the
curve of $\delta_a (E)$ is characterized by the line profile index
$q_a = -2 \Delta\Gamma / ( 2\sqrt{\Gamma_1 \Gamma_2 } \cos
\Delta_{12}^{0} )$. The strength of the avoided crossing interaction
felt by the eigenphase shifts is governed by the magnitude of
$\Gamma_a$. No shift is expected 
when $\Gamma_1$ equals $\Gamma_2$, where the strength of the avoided
crossing interaction is  strongest. The maximum shift obtains when
either $\Gamma_1$ or $\Gamma_2$ is zero. 

Eigenvectors $v_{\pm}$ of $A$ of $S=S^{0}A$ corresponding to
eigenvalues $\delta_{\pm} (E)$ are obtained as
\[
v_{+} = \left[ \begin{array}{c}
\cos (\frac{\theta_{a}}{2}) \\
 \sin (\frac{\theta_{a}}{2})  
\end{array} \right] ,  ~~
v_{-} = \left[ \begin{array}{c}
- \sin (\frac{\theta_{a}}{2}) \\
\cos (\frac{\theta_{a}}{2}) 
\end{array} \right] , 
\]
where $\theta_{a}$ is defined by
\[
\cos\theta_{a} = -\frac{\epsilon_a}{\sqrt{1+\epsilon_a
    ^2}} , ~~
\sin\theta_{a} = \frac{1}{\sqrt{1+\epsilon_a
    ^2}} .
\]
Eigenvectors are independent of $q_{a}$. They
depend only on $\epsilon_a$. Since $q_{a}$
stands for the apartness of 
the avoided crossing point energy from the middle of asymptotes of the
abscissas of two
eigenphase shifts, the above result means that the characteristics
of avoided crossing interactions are independent of the asymptotes of
the abscissas  of eigenphase shifts. The corresponding eigenvectors of
$S$ matrix are 
obtained by replacing $\theta_a$ with $\theta_a' = \theta_a + 2
\theta_{0}$. 

With the new parameters, $S$ matrix is found to be
expressible as
\[
S = e^{-i\left(  \delta_{\Sigma} +\delta_a 
\vec{\sigma} \cdot \hat{n}_{\theta_a'}  \right) } ,
\] 
where $\hat{n}_{\theta_a'} = \hat{z} \cos \theta_a' + \hat{x} \sin
\theta_a'$. 
In terms of Pauli matrices, the time delay matrix $Q = i\hbar
S^{\dag}\frac{dS}{dE}$ is found to be consisted of three terms:
\[
Q = \frac{1}{2} (\tau_r + 
\vec{\sigma}\cdot \hat{n}_{\theta_a'} \tau_a 
+ \vec{\sigma}\cdot \hat{n}_{\theta_a'}^{\perp} \tau_f ) ,
\]
the one due to the resonance, the one due to the avoided crossing
interaction, 
and the one due to the change of frame as a function of
energy. Because of the last term, eigenfunctions of $Q$ matrix are
different from those of $S$ matrix.
 
The time delay due to the 
resonance takes a symmetric Lorentzian form and the time delay due the
to avoided 
crossing takes a form of a functional of the Beutler-Fano formula:
\[
\tau_a (E) = 
\left\{ \begin{array}{ll}
- \tau_r (E) 
\frac{1}{\sqrt{1+r^2 \frac{1+\epsilon_r^2}{(\epsilon_r - q_{\tau}
      )^2}}},&{\rm when}~\epsilon_r \le q_{\tau} ,\\
 \tau_r (E) 
\frac{1}{\sqrt{1+r^2 \frac{1+\epsilon_r^2}{(\epsilon_r - q_{\tau}
      )^2}}},&{\rm when}~\epsilon_r > q_{\tau} .
\end{array} \right.
\]
The asymmetry of $\tau_a$ as a function of energy is brought about by
the nonzero value of $q_{\tau}$ which is proportional to the shift of
the avoided crossing point energy from the resonance one. Thus the
asymmetry of $\tau_a$ is caused by the mismatch in the positions of
the avoided crossing point and resonance  energies.

The time delay due to frame change takes the following form:
\[
\tau_f (E) = \tau_r (E) \sin \Delta_{12}^{0}
\sqrt{\frac{\epsilon_r^2 +1}{\epsilon_a^2 +1} }
. \] 
The above form may be understood from the fact that the changes of the
frame spanned by the eigenvectors of $S$ matrix is governed by
$1+ \epsilon_a ^2$ while the time delay due to the resonance is
governed 
by $1+ \epsilon_r^2$. Therefore the ratio of $\tau_a$ to $\tau_r$ will be
a function of $(1+ \epsilon_r^2 )/(1+ \epsilon_a^2 )$.  $\tau_f
(E)$ is also found to be transformed into a functional  of the
Beutler-Fano formula. 

In analogy with the spin $\frac{1}{2}$ system, the time delay matrix
$Q$ may be expressed in terms of polarization vectors and the Pauli
spin matrices as
\[
Q = \frac{1}{2} \tau_r \left( 1 + \vec{P}_a \cdot \vec{\sigma} + \vec{P}_f
\cdot \vec{\sigma} \right) ,
\]
where 
polarization vectors are defined by
\[
\vec{P_a} = \frac{\tau_a}{\tau_r} \hat{n}_{\theta_a '} , ~~~ 
\vec{P_f} = \frac{\tau_f}{\tau_r} \hat{n}_{\theta_a '}^{\perp}
. 
\]
Like the spin $\frac{1}{2}$ system, it is found that the absolute
values of $\vec{P}_a$ and $\vec{P}_f$ are restricted to $0 \le
|\vec{P}_a | \le 1$ and $0 \le | \vec{P}_f | \le 1$. In the present
case a complete depolarization means that eigentimes delays are the
same 
regardless of eigenchannels, while a complete polarization means that
eigentime delays are 0 and $\tau_r (E )$ as a function of energy.
Eigenvectors for eigentime delays due to an avoided crossing
interaction and due to a frame 
change are orthogonal to each other and contribute to
the total eigentime delays as $\sqrt{\tau_a ^2 + \tau_f ^2} = \tau_r
\sqrt{P_a ^2 + P_f^2}$. It is found that $P_a^2 + P_f^2
=1$. This means that one among two total eigentime delays is zero
while the other one is the same as the time delayed by the resonance
state. Though time delays due to an avoided crossing interaction and
a frame 
change are asymmetric with respect to the resonance energy and
therefore the energy of the longest lifetime not matched with the
resonance energy, the energy of the longest overall eigentimes delayed
is exactly matched with the resonance energy. Though this is a
surprising result, it should rather be so if we recall that
all the partial life times obtained from the partial photo-dissociation
cross 
sections as a function of energy are the same in the system of one
discrete state and many continua. But the detailed study on their
connection is beyond the scope of this paper.

Present work reveals the dynamical parameters that govern the
behaviors of eigenphase shifts and eigentime delays for the system of
one discrete state and two continua. It may be applied to the systems
of more than two continua by approximating that such systems are
a cascade of independent two interacting continua. In that sense,
current study may be considered as a prototype model for the
``isolated'' avoided crossing interaction.  For the system of one
discrete state and 
three or four continua, Eq. (\ref{Sardi}) becomes third and fourth
order equation and its solution can be obtained by the Cartan and
Ferrari's method. It may be highly desirable to do the similar studies
on these systems. 

\acknowledgements I would like to thank to Professors Fano and Nakamura
for their interests and advice on this work. Careful readings and
comments by Young-Man Han are also appreciated. This work was supported
by KOSEF under contract No. 961-0305-050-2 and at the end stage by
Korean Ministry of 
Education through Research Fund No. BSRI-97-3449.

\pagebreak

\begin{figure}
\epsfig{file=eigen.eps,width=13cm}
\caption{$\tilde{\delta}_a (E)$ vs. $\epsilon_a$ and
$\bar{\delta}_{\pm} (\epsilon_r )$
  vs. $\epsilon_r$ are plotted for $q_{a}$ = 0, $\pm 1$, and $\pm
  3$. Values for $\delta_1^{0}$ and 
  $\delta_2^{0}$ are $\pi/3$ and $\pi/6$, respectively.~~~~~~~~~~~~~~~~~~~~~~~~
	}
\label{fig:delta}
\end{figure}

\begin{figure}
\epsfig{file=g_dab_df.eps,width=11cm}
\caption{$g_a (\epsilon_r)$ and $\bar{\tau}_a (\epsilon_r )$
  vs. $\epsilon_r$ 
are plotted for three different profile indices $q_{\tau}$ = 0.6, 1, and
5 with $\Delta_{12}^{0}$ = $\pi/3$.~~~~~~~~~~~~~~~~~~~~~~~~~~~~~~~~~
} 
\label{fig:g_td_a}
\end{figure}

\begin{references}
\bibitem{13} With the choice of the sign for $\delta_a$ as in
Eq. (\ref{avoiding_term}), we obtain the following formulas:
\[
\sin \delta_a = \frac{\Gamma_a}{\Gamma} \sin \Delta_{12}^{0}
\sqrt{\frac{1+ \epsilon_a^2}{1+\epsilon_r^2}} , ~~~ 
\cos \delta_a = - \frac{\Gamma_a}{\Gamma} \cos \Delta_{12}^{0}
\frac{\epsilon_a - q_a}{\sqrt{1+\epsilon_r^2}} ,
\] 
whose forms are consistent with those for $\theta_a$ and $\delta_r$
defined in Eqs. (\ref{trig_theta_a}) and (\ref{delta_single}).
\end{references}
\end{document}